%% file: eta23pi.tex
\documentclass[prd,amsmath,amssymb,showpacs,superscriptaddress,nofootinbib,twocolumn]{revtex4-1}
\usepackage{graphicx}
\usepackage{epsfig,graphics,subfigure,psfrag,amsmath,amssymb}
\usepackage{amsmath}
\usepackage[mathlines]{lineno}
\usepackage{dcolumn}
\usepackage{color}
\usepackage{bm}
\usepackage{overpic}
\newcommand{\ra}{\rightarrow}

\newcommand{\jpsi}{J/\psi}
\newcommand{\pio}{\pi^{0}}
\newcommand{\pip}{\pi^{+}}
\newcommand{\pim}{\pi^{-}}
\newcommand{\etap}{\eta^{\prime}}
\newcommand{\chisq}{\chi^{2}}
\renewcommand{\thefootnote}{\fnsymbol{footnote}}

\usepackage[ 
total={6.5in,8.75in}, top=1.2in, left=0.9in, 
]{geometry}

\begin{document}

\title{\boldmath Measurement of the Matrix Elements for the Decays $\eta \ra \pi^{+}\pi^{-}\pi^0$ and
$\eta/\etap\ra\pi^0\pi^0\pi^0$}

\input{authors_jun2015}

\begin{abstract}
Based on a sample of $1.31\times 10^9$ $J/\psi$ events collected with the BESIII detector at the BEPCII collider,
Dalitz plot analyses of selected 79,625 $\eta\ra\pi^{+}\pi^{-}\pi^0$ events, 33,908 $\eta\ra\pi^0\pi^0\pi^0$ events and 1,888
$\etap\ra\pi^0\pi^0\pi^0$ events are performed.
The measured matrix elements of
$\eta\ra\pi^+\pi^-\pi^0$ are in reasonable agreement with previous
measurements. The Dalitz plot slope parameters of $\eta\ra\pi^0\pi^0\pi^0$
and $\etap\ra\pi^0\pi^0\pi^0$ are determined to be $-0.055\pm 0.014\pm0.004$
and $-0.640\pm 0.046\pm0.047$, respectively, where the first uncertainties are statistical
and the second systematic. Both values are consistent with
previous measurements, while the precision of the latter one is
improved by a factor of three. Final state interactions are found to have an important role in those decays.
\end{abstract}

\maketitle

\section{INTRODUCTION}

Since the electromagnetic contribution to the isospin violating decays $\eta/\eta^\prime\ra3\pi$
is strongly suppressed~\cite{Bell1968315,Baur1996127,Ditsche200983},
the decays are
induced dominantly by the strong interaction.
Therefore, they offer a unique opportunity to investigate
fundamental symmetries and measure the $u-d$ quark mass difference.
At the tree level of chiral
perturbation theory (ChPT), the predicted decay width of
$\eta\rightarrow\pi^+\pi^-\pi^0$~\cite{Osborn1970} is about 70 eV, which is much lower than the experimental
value of $300\pm11$ eV~\cite{PDGgroup}. To explain this discrepancy, considerable theoretical effort
has been made, including a
dispersive approach~\cite{Anisovich1996335} and non-relativistic effective field theory~\cite{NREFT}.
Recently, it was found that
higher order terms in ChPT at
next-to leading order (NLO)~\cite{Gasser1985539}
and next-next-to leading order (NNLO)~\cite{Bijnens:2007pr}
are crucial for a comparison with experimental results, where $\pi\pi$
re-scattering between the final
state pions is present.

To distinguish between the different theoretical approaches, precise
measurements of the matrix elements for $\eta\rightarrow\pi^+\pi^-\pi^0$ and the decay width are important.
For the three-body decay $\eta\rightarrow\pi^+\pi^-\pi^0$, the decay amplitude square can be parameterized as~\cite{Ambrosino:2008ht}
\begin{equation}\label{eq:etacha_amp}
\begin{gathered}
        |A(X,Y)|^{2} = N(1 + aY + bY^2 + cX + dX^2 \\
        						     + eXY + fY^{3} + \ldots),
\end{gathered}
\end{equation}
where $X$ and $Y$ are the two independent Dalitz plot variables defined as
\begin{equation}\label{eq:xy}
\begin{gathered}
    X=\frac{\sqrt{3}}{Q}(T_{\pi^{+}}-T_{\pi^{-}}),\\
    Y=\frac{3T_{\pi^{0}}}{Q}-1,
\end{gathered}
\end{equation}
where $T_{\pi}$ denotes the kinetic energy of a given pion in the $\eta$ rest frame,
$Q=m_{\eta}-m_{\pip}-m_{\pim}-m_{\pio}$ is the excess energy of the reaction,
$m_{\eta/\pi}$ are the nominal masses from PDG~\cite{PDGgroup},
and $N$ is a normalization
factor. The coefficients $a, b, c, \ldots$ are the Dalitz plot parameters,
which are used to test theoretical
predictions and fundamental symmetries. For example, a non-zero value for the odd powers of $X$, $c$ and $e$,
implies the violation of charge conjugation.

The Dalitz plot distribution of $\eta\rightarrow\pi^{+}\pi^{-}\pi^0$ has been analyzed previously
by various experiments~\cite{PDGgroup}.
Using a data sample corresponding to about $5\times 10^{6}$ $\eta$ mesons produced in
$e^+e^-\ra\phi\ra\gamma\eta$ reactions, KLOE~\cite{Ambrosino:2008ht}
provided the most precise measurement, where
the Dalitz plot parameters $c$ and $e$
are found to be consistent with zero within uncertainties, and $f$ was measured for the first time. Most
recently, the WASA-at-COSY collaboration analyzed $\eta\rightarrow\pi^+\pi^-\pi^0$ based on a data sample
corresponding to $1.2\times10^7$ $\eta$ mesons produced in $pd$ $\ra ^3$He $\eta$ reactions at 1 GeV~\cite{PhysRevC.90.045207}.
The results are in agreement with those from KLOE within
two standard deviations.

For $\eta/\etap\ra\pio\pio\pio$, the density distribution of the Dalitz plot has threefold symmetry
due to the three identical particles in the final state.
Hence, the density distribution can be parameterized using polar variables~\cite{Unverzagt2009169}
\begin{equation}\label{zdef}
    Z=X^2+Y^2=\frac{2}{3}\sum^{3}_{i=1}(\frac{3T_{i}}{Q}-1)^{2},
\end{equation}
and the expansion
\begin{equation}\label{eq:etaneu_amp}
    |A(Z)|^2=N(1 + 2\alpha Z + \ldots),
\end{equation}
where $\alpha$ is the slope parameter,
$Q = m_{\eta/\etap}-3m_{\pio}$,
$T_{i}$ denotes the kinetic energies of each $\pio$ in the $\eta/\etap$ rest frame and
$N$ is a normalization factor.
A non-zero $\alpha$ indicates final-state interactions.

The world averaged value of the Dalitz plot slope parameter $\alpha=-0.0315\pm0.0015$~\cite{PDGgroup} for
$\eta\rightarrow \pi^0\pi^0\pi^0$ is dominated by the measurements of the
Crystal Ball~\cite{Unverzagt2009169}, WASA-at-COSY~\cite{Adolph200924}
and KLOE~\cite{Ambrosino201016} experiments.
Interestingly, the predicted value for $\alpha$ in NLO and NNLO
ChPT~\cite{ref:etaneuwithcor,ref:Beisert2003186,Bijnens:2007pr} is positive,
although the theoretical uncertainties are quite large.

The decay $\eta^\prime\rightarrow \pi^0\pi^0\pi^0$ has been explored with very limited statistics only.
The GAMS-2000 experiment reported
the first observation of $\eta^\prime\rightarrow \pi^0\pi^0\pi^0$~\cite{ref:etapneugam2} and measured
the Dalitz plot slope with 62 reconstructed events. This result was
later updated to be $\alpha=-0.59\pm0.18$~\cite{ref:etapneugam}
with 235 events. In 2012, the same decay was investigated by BESIII~\cite{Ablikim2012182001}
using a data sample of $225\times 10^6$ $\jpsi$ events.
The branching fraction was measured to be about twice as large as the previous measurements,
but the Dalitz plot slope parameter was not measured.

In this paper, the matrix elements for $\eta\ra\pi^{+}\pi^{-}\pi^0$ and
$\eta/\etap\ra\pio\pio\pio$ are measured, 
where the Dalitz plot slope parameter of $\etap\ra\pio\pio\pio$ is determined
with higher precision than the existing measurements.
This analysis is performed using a sample of $1.31\times 10^9$ $\jpsi$ events accumulated with the BESIII detector.
Radiative $\jpsi\ra\gamma\eta^{(\prime)}$ decays are exploited to access the $\eta$ and $\etap$ mesons.

\section{DETECTOR AND MONTE CARLO SIMULATION}

BEPCII is a double-ring $e^{+}e^{-}$ collider working
at center-of-mass energies from 2.0 to 4.6 GeV.
The BES\uppercase\expandafter{\romannumeral3}~\cite{Ablikim2010345} detector at BEPCII collider, with a geometrical
acceptance of 93\% of 4$\pi$ stereo angle, operates in a 1.0 T (0.9 T in 2012, 
when about 83\% of the data sample were collected) 
magnetic field provided by a
superconducting solenoid magnet. The detector is composed of a helium-based drift chamber (MDC),
a plastic-scintillator time-of-flight (TOF) system, a CsI(Tl) electromagnetic calorimeter (EMC)
and a multi-layer resistive plate counter system (MUC). The charged-particle momentum
resolution at 1.0 GeV/$c$ is 0.5\%, and the specific energy loss ($dE/dx$) resolution is better than 6\%. The spatial
resolution of the MDC is better than 130 $\mu$m. The time resolution of the TOF is 80 ps in the barrel and
110 ps in the endcaps. The energy resolution of the EMC at 1.0 GeV/$c$ is 2.5\% (5\%) in the barrel (endcaps),
and the position resolution is better than 6 mm (9 mm) in the barrel (endcaps).
The position resolution in the MUC is better than 2 cm.

Monte Carlo (MC) simulations are used to estimate backgrounds and determine the detection efficiencies.
The {\sc GEANT}4-based~\cite{ref:geant4} simulation software {\sc BOOST}~\cite{ref:boost} includes the geometric and material
description of the BESIII detector, detector response, and digitization models,
as well as the tracking of the detector running conditions and performance.
The production of the $\jpsi$ resonance is simulated with
{\sc KKMC}~\cite{ref:kkmc,ref:kkmc2}, while the decays are generated with {\sc EVTGEN}~\cite{ref:evtgen}
for known decay modes with branching fractions being set to the world average values~\cite{PDGgroup}
and by {\sc LUNDCHARM}~\cite{ref:lundcharm} for the remaining unknown decays.
We use a sample of $1.2 \times 10^9$ simulated $J/\psi$ events where the $J/\psi$ decays generically (`inclusive MC
sample') to identify background contributions.
The analysis is performed in the framework of the BESIII offline software system
({\sc BOSS})~\cite{ref:boss} which takes care of the detector calibration, event reconstruction, and data storage.

\section{MEASUREMENT OF THE MATRIX ELEMENTS FOR THE DECAY $\eta\ra\pip\pim\pio$}

For the reconstruction of $\jpsi\ra\gamma\eta$ with $\eta\ra\pip\pim\pio$ and $\pio\ra\gamma\gamma$,
events consistent with the topology $\pip\pim\gamma\gamma\gamma$ are selected and the following criteria are applied.
For each candidate event, we require that two charged tracks are reconstructed in the MDC
and the polar angles of the tracks satisfy $|\cos\theta| < 0.93$. The tracks are required to pass the interaction point within
$\pm10$ cm along the beam direction and
within $1$ cm in the plane perpendicular to the beam.
Photon candidates are reconstructed using clusters of energy deposited in the EMC.
The energy deposited in nearby TOF counters is included in EMC measurements to improve the reconstruction efficiency
and the energy resolution. Photon candidates are required to have a deposited energy larger than 25 MeV in
the barrel region ($|\cos\theta|<0.80$) and 50 MeV in the endcap region ($0.86<|\cos\theta|<0.92$).
In order to eliminate clusters associated with charged tracks, the angle between the directions of any charged track and the photon candidate
must be larger than 10$^\circ$. 
Requirements of EMC cluster timing with respect to the event start time are used
to suppress electronic noise and energy deposits unrelated to the event.
Events with exactly two charged tracks
of opposite charge and at least three photon candidates that satisfy the above requirements are retained for further analysis.

The photon candidate with the largest energy in the event is regarded as the radiative photon originating
from the $\jpsi$ decays. For each $\pip\pim\gamma\gamma\gamma$ combination, a six constraints (6C) kinematic fit is performed.
The fit enforces energy-momentum conservation, and the invariant masses of $\gamma\gamma$ and $\pip\pim\pio$
are constrained to the nominal $\pio$ and $\eta$ mass, respectively.
Events with a $\chisq$ from the 6C-kinematic fit ($\chisq_{6C}$) less than 80 are accepted for further analysis.
If there are more than three photon candidates in an event, only the combination with the smallest $\chisq_{6C}$
is retained. To reject possible backgrounds with two or four photons in the final state, kinematic fits are also performed with
four constraints enforcing energy-momentum conservation under the $\jpsi\ra\pip\pim\gamma\gamma\gamma$ signal hypothesis
as well as the $\jpsi\ra\pip\pim\gamma\gamma\gamma\gamma$ and $\jpsi\ra\pip\pim\gamma\gamma$ background hypotheses. 
Events with a $\chisq_{4C}$ value for the signal hypothesis greater than
that of the $\chisq_{4C}$ for any background hypothesis are discarded.

\begin{figure}
		\includegraphics[width=0.45\textwidth]{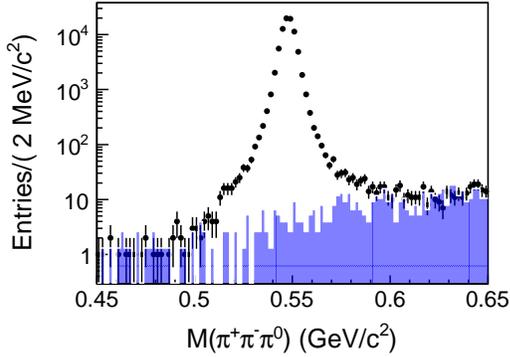}
    \caption{\label{fig:M3pi_cha} Invariant mass spectrum of $\pip\pim\pio$ obtained after the kinematic fit
		without the
$\eta$ mass constraint applied. The dots with error bars are for data and the shaded histogram
is for background events estimated from the inclusive MC sample. }
\end{figure}

After applying the selection criteria described above, 79,625 $\eta\rightarrow \pi^+\pi^-\pi^0$
candidate events are selected. To estimate the background
contribution under the $\eta$ peak, we perform an alternative selection, where the
$\eta$ mass constraint in the kinematic fit is removed. The resulting invariant mass spectrum of $\pi^+\pi^-\pi^0$,
$M(\pip\pim\pio)$, is shown in Fig.~\ref{fig:M3pi_cha}. A significant $\eta$ signal is observed
with a low background level.
The background contamination is estimated to be 0.2\% from $\eta$ sideband regions,
defined as $0.49<M(\pip\pim\pio)<0.51$ GeV/$c^2$ and $0.59<M(\pip\pim\pio)<0.61$ GeV/$c^2$, in the data sample.
In addition, a sample of $1.2 \times 10^9$ inclusive MC $J/\psi$ decays is used to investigate potential backgrounds.
Using the same selection criteria, the distribution of $M(\pip\pim\pio)$ for this sample
is depicted as the shaded histogram in Fig.~\ref{fig:M3pi_cha}.
No peaking background remains around the $\eta$ signal region.
The background contamination is estimated to be about 0.1\%, and is therefore
not considered in the extraction of the Dalitz plot parameters.

\begin{figure}[htbp]
\includegraphics[width=0.45\textwidth,height=6cm]{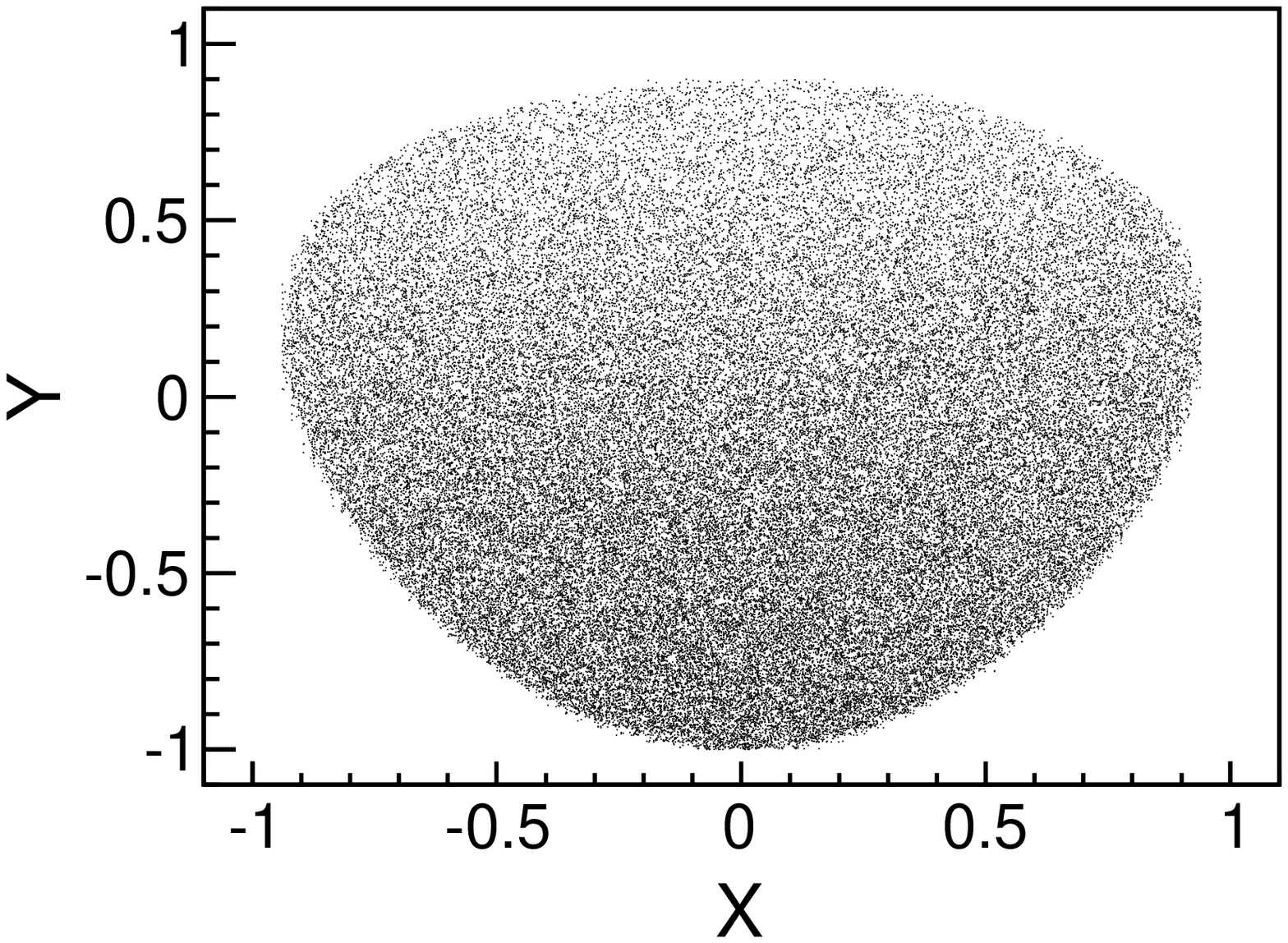}
    \caption{ \label{fig:etacha_dalXY} Dalitz plot for $\eta\ra\pip\pim\pio$ in the data sample.}
\end{figure}

\begin{figure*}[htbp]
\includegraphics[width=0.45\textwidth]{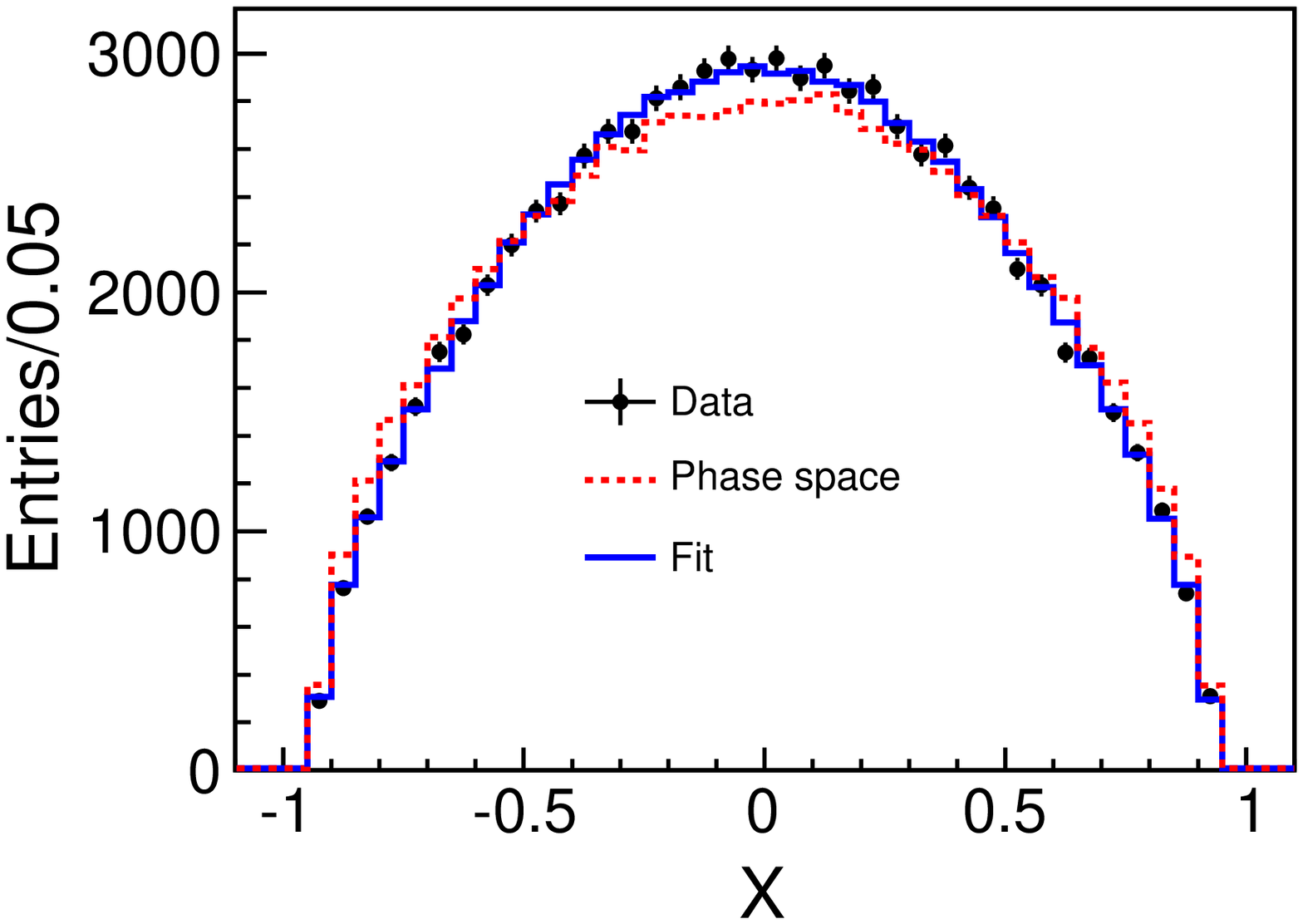}\put(-45,120){\bf (a)}
\includegraphics[width=0.45\textwidth]{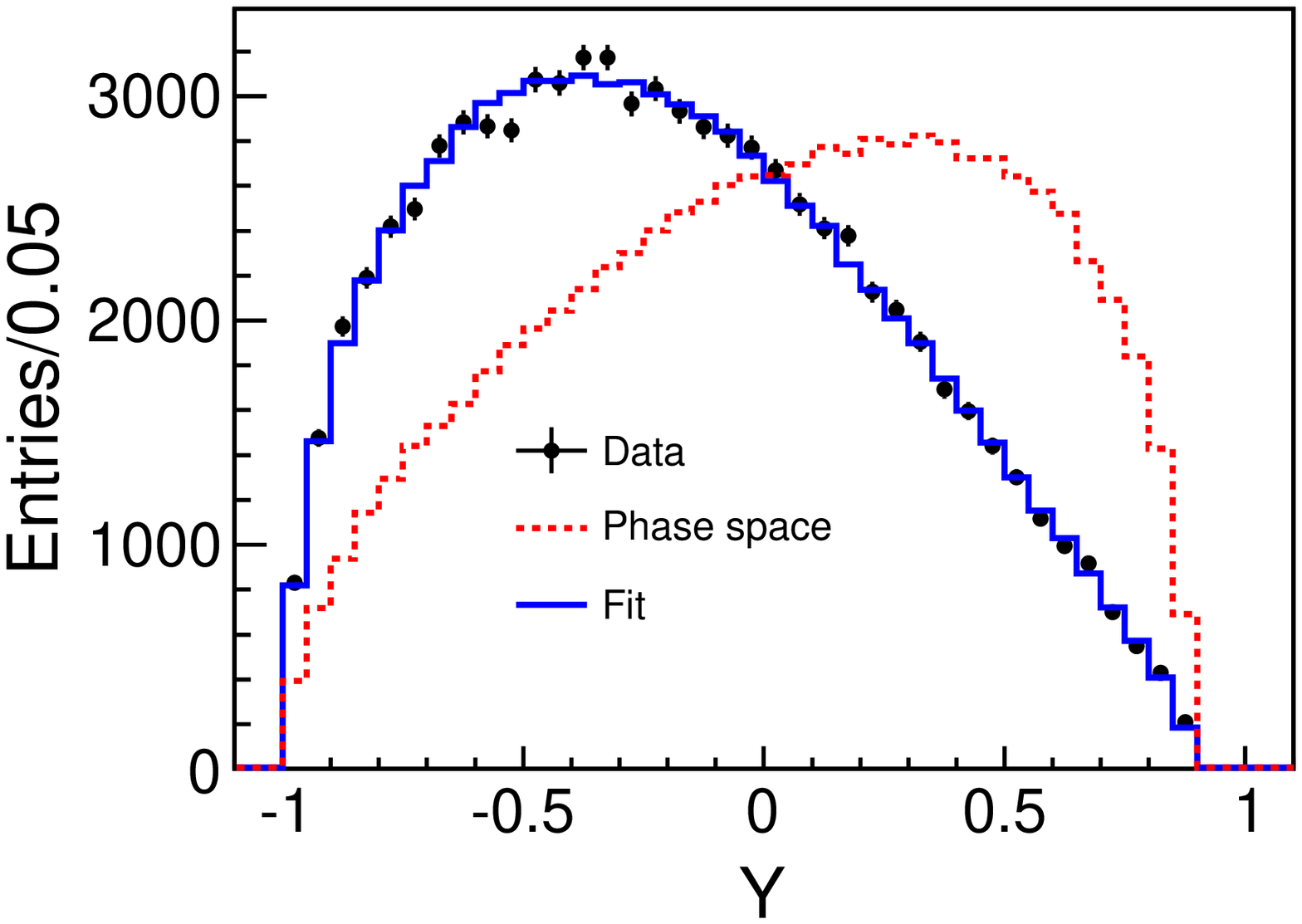}\put(-45,120){\bf (b)}
    \caption{\label{fig:etacha_dalXY_fit} Projections of the Dalitz plot (a) $X$ and (b) $Y$
			for $\eta\ra\pip\pim\pio$ obtained from data (dots with error bars)
			and phase space distributed MC events (dashed line).
			The result of the fit described in the text (solid line) is also plotted.}
\end{figure*}

The Dalitz plot in the variables $X$ and $Y$ is shown
in Fig.~\ref{fig:etacha_dalXY} for the selected events.
The $X$ and $Y$ projections are shown in Fig.~\ref{fig:etacha_dalXY_fit}. 
For comparison, the corresponding distributions obtained from MC events with
phase space distributed $\eta\ra\pip\pim\pio$ decays are also shown. 
The phase space MC distributions of $X$ and $Y$ differ visibly from those in the data sample, which indicates 
there could be large contributions from higher order terms in ChPT.

In order to investigate the dynamics of $\eta\ra\pip\pim\pio$, the Dalitz plot matrix elements of the decay amplitude
given in Eq.~(\ref{eq:etacha_amp}) are obtained from an unbinned maximum likelihood fit to the data.
To account for the resolution and detection efficiency, the amplitude is convoluted with a function
$\sigma(X,Y)$ parameterizing the resolution, and multiplied by a function $\varepsilon(X,Y)$
parameterizing the detection efficiency. Both functions are derived from MC simulations.
The sum of two Gaussian functions is used for $\sigma(X,Y)$,
while $\varepsilon(X,Y)$ is a quadratic function. After normalization, one derives the probability density
function ${\cal P}(X,Y)$, which is applied in the fit:
\begin{equation}
\begin{gathered}
	{\cal {P}}(X,Y) = \\
	\frac{(|A(X,Y)|^{2}\otimes\sigma(X,Y))\varepsilon(X,Y)}{\int_{DP}{(|A(X,Y)|^{2}\otimes\sigma(X,Y))\varepsilon(X,Y)dXdY}},
\end{gathered}
\end{equation}
where $A(X,Y)$ is the decay amplitude of $\eta\ra\pip\pim\pio$ and the integral taken over the Dalitz plot
(DP) accounts for normalization.

For the fit, the negative log-likelihood value
\begin{equation}
	-\ln {\cal L} = -\sum_{i=1}^{N_\text{event}}\ln{}{{\cal{P}}(X_{i},Y_{i})}
\end{equation}
is minimized, where ${\cal P}(X_{i},Y_{i})$ is evaluated for each event $i$, 
and the sum includes all accepted events.

We perform two fits to the data. For the first fit, we assume charge conjugation invariance and we fit
the parameters for the matrix elements $a, b, d$ and $f$ only, while $c$ and $e$ are set to zero.
For the second fit, we include the possibility of charge conjugation violation and the latter two parameters are also allowed to vary
in the fit.

In the case of charge conjugation invariance, the fit yields the following parameters (with statistical errors only)
\begin{equation}
\begin{matrix}
a & = & -1.128&\pm&0.015, \\
b & = & \phantom{-} 0.153 &\pm&0.017, \\
d & = & \phantom{-} 0.085 &\pm&0.016,  \\
f & = & \phantom{-} 0.173 &\pm&0.028. \\
\end{matrix}
\end{equation}
The corresponding correlation matrix of the fit parameters is given by
\begin{equation}
\begin{pmatrix}
   & \vline & b & d & f\\\hline
 a & \vline & -0.265 & -0.389 & -0.749\\
 b & \vline & \phantom{-}  1.000 & \phantom{-}  0.311 & -0.300\\
 d & \vline &        &  \phantom{-} 1.000 &\phantom{-}  0.079\\
\end{pmatrix}
.
\end{equation}

The fit projections on $X$ and $Y$, illustrated as the solid histograms in Fig.~\ref{fig:etacha_dalXY_fit},
indicate that the fit can describe the data well. The obtained parameters are
in agreement with previous measurements within two standard deviations.

If the possibility of charge conjugation violation is included in the decay amplitude,
the fit to data yields the following results (with statistical uncertainties only)
\begin{equation}
\begin{matrix}
a & = &  -1.128 &  \pm & 0.015, \\
b & = & \phantom{-}  0.153  &  \pm & 0.017, \\
c & = &  (0.047 &  \pm & 0.851)\times10^{-2}, \\
d & = & \phantom{-}  0.085  &  \pm & 0.016, \\
e & = & \phantom{-}  0.017  &  \pm & 0.019, \\
f & = & \phantom{-}  0.173  &  \pm & 0.028.
\end{matrix}
\end{equation}
The corresponding correlation matrix of the fit parameters is given by
\begin{widetext}
\begin{equation}
\begin{pmatrix}
   		& \vline &  b 		 & c      &   d    &  e    &  f\\\hline
a			& \vline &  -0.265 & -0.003 & -0.388 & \phantom{-} 0.001 & -0.749 \\
b			& \vline &  \phantom{-}  1.000 & -0.001 & \phantom{-}  0.311 & \phantom{-} 0.016 & -0.300 \\
c			& \vline &         &  \phantom{-} 1.000 &  \phantom{-} 0.003 &-0.592 &  \phantom{-} 0.003 \\
d			& \vline &         &        &  \phantom{-} 1.000 & \phantom{-} 0.016 &  \phantom{-} 0.079 \\
e			& \vline &         &        &        & \phantom{-} 1.000 & -0.007 \\
\end{pmatrix}
.
\end{equation}
\end{widetext}

Compared with the fit results assuming charge-parity conservation, the derived parameters $a$, $b$, $d$ and $f$ are almost unchanged.
The parameters $c$ and $e$ are consistent with zero within one standard deviation, which indicates that
there is no significant charge-parity violation in decay $\eta\ra\pip\pim\pio$.
Comparing the two fits, the significance of charge-parity violation is determined to be only 0.65$\sigma$.

The fit procedure is verified with MC events that were generated based on the Dalitz plot matrix elements from the fit to the data.
Following the same reconstruction and fitting procedure as applied to the data sample,
the extracted values are consistent with the input values of the simulation.

\section{MEASUREMENT OF THE MATRIX ELEMENT FOR THE DECAYS $\eta\ra\pio\pio\pio$ and $\etap\ra\pio\pio\pio$ }

For the reconstruction of $\jpsi\ra\gamma\eta/\etap$ with $\eta/\etap\ra\pio\pio\pio$ and $\pio\ra\gamma\gamma$,
events containing at least seven photon candidates and no charged tracks are selected.
The selection criteria for photons are the same
as those described above for $\eta\ra\pip\pim\pio$, 
except the requirement of the angle between the photon candidates and any charged track.
Requirements of EMC cluster timing with respect to the most energetic photon are also used.
Again, the photon with the largest energy in the event is assumed to be
the radiative photon originating from the $\jpsi$ decay. From the remaining candidates, pairs of photon are combined
into $\pio\ra\gamma\gamma$ candidates which are subjected to a kinematic fit,
where the invariant mass of the photon pair is constrained
to the nominal $\pio$ mass. The $\chisq$ value of this kinematic fit with one degree of freedom
is required to be less than 25.
To suppress the $\pio$ mis-combination, the $\pio$ decay angle $\theta_{\rm decay}$, defined as the polar angle of a photon in the
corresponding $\gamma\gamma$ rest frame, is required to satisfy $|\cos \theta_{\rm decay}|<0.95$. 
From the accepted $\pio$ candidates and the corresponding radiative photon,
$\gamma\pio\pio\pio$ combinations are formed.
A kinematic fit with seven constraints (7C) is performed,
enforcing energy conservation and constraining the invariant mass of $\gamma\gamma$ pairs to the
nominal $\pio$ mass. 
If more than one combination is found in an event, only the one with the smallest $\chisq_{7C}$ is retained.
Events with $\chisq_{7C} < 70$ are accepted for further analysis.

For $\etap\ra\pio\pio\pio$, backgrounds from $\jpsi\ra\omega\pio\pio$ 
are suppressed by vetoing events with
$|M(\gamma\pio)-m_{\omega}|<0.05$ GeV/$c^2$, where $M(\gamma\pio)$ is the invariant mass of the $\gamma\pio$ combination
closest to the nominal $\omega$ mass ($m_{\omega}$)~\cite{PDGgroup}.
Peaking backgrounds for the process $\etap\ra\pio\pio\pio$ can arise from $\jpsi\ra\gamma\etap$ with $\etap\ra\eta\pio\pio$.
To suppress these backgrounds, a 7C kinematic fit under the $\jpsi\ra\gamma\eta\pio\pio$ hypothesis is performed.
Events for which the $\chisq$ value obtained for the background hypothesis is less than that obtained
for the $\gamma\pio\pio\pio$ hypothesis are discarded.
In addition, events with an invariant mass of at least one $\gamma\gamma$ pair in the mass window
$|M(\gamma\gamma)-m_{\eta}|<0.03$ GeV/$c^2$
are rejected.

\begin{figure*}[!htbp]
		\includegraphics[width=8cm, height=6cm]{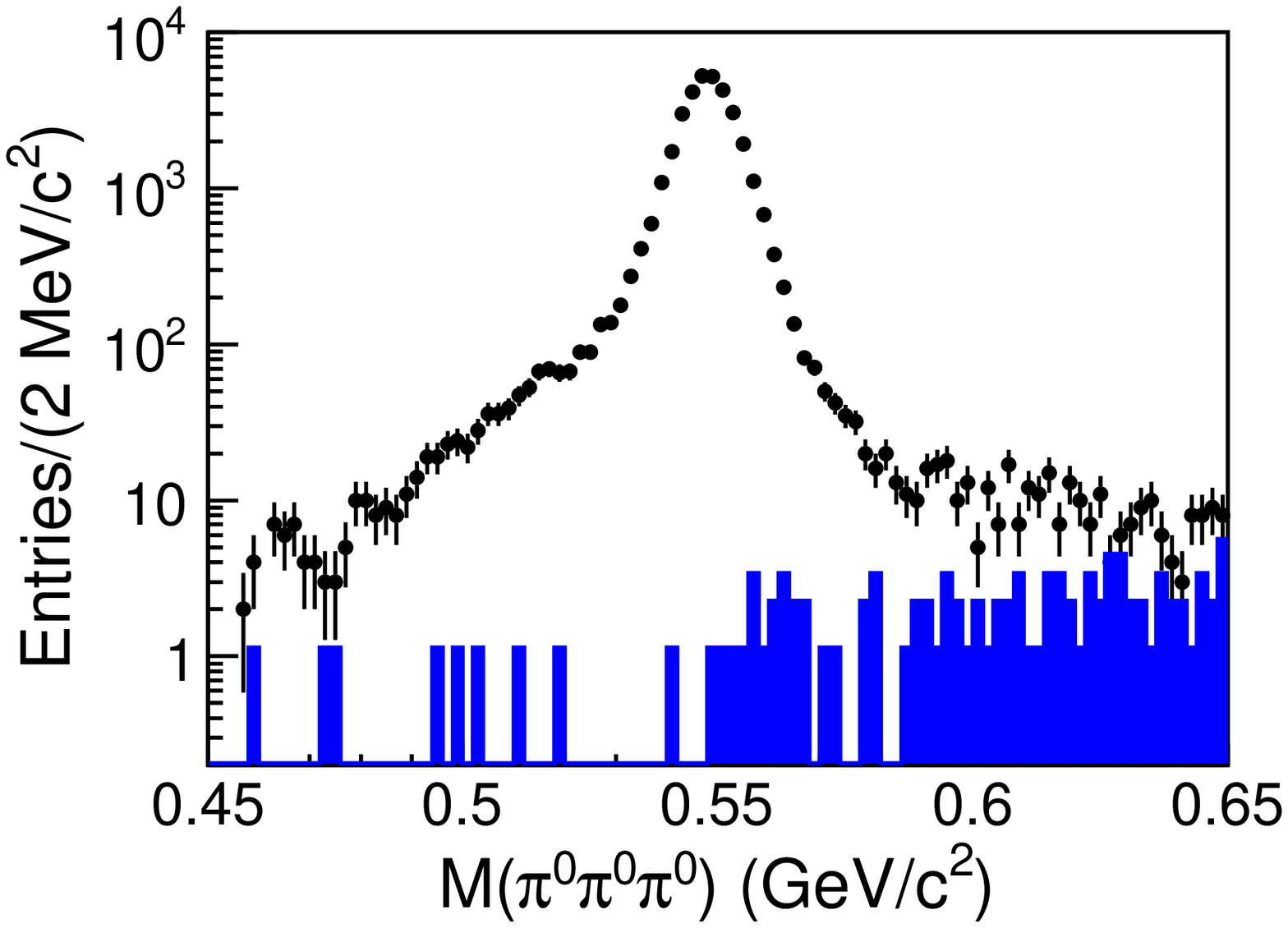}\put(-45,140){\bf (a)}
    \includegraphics[width=8cm, height=6cm]{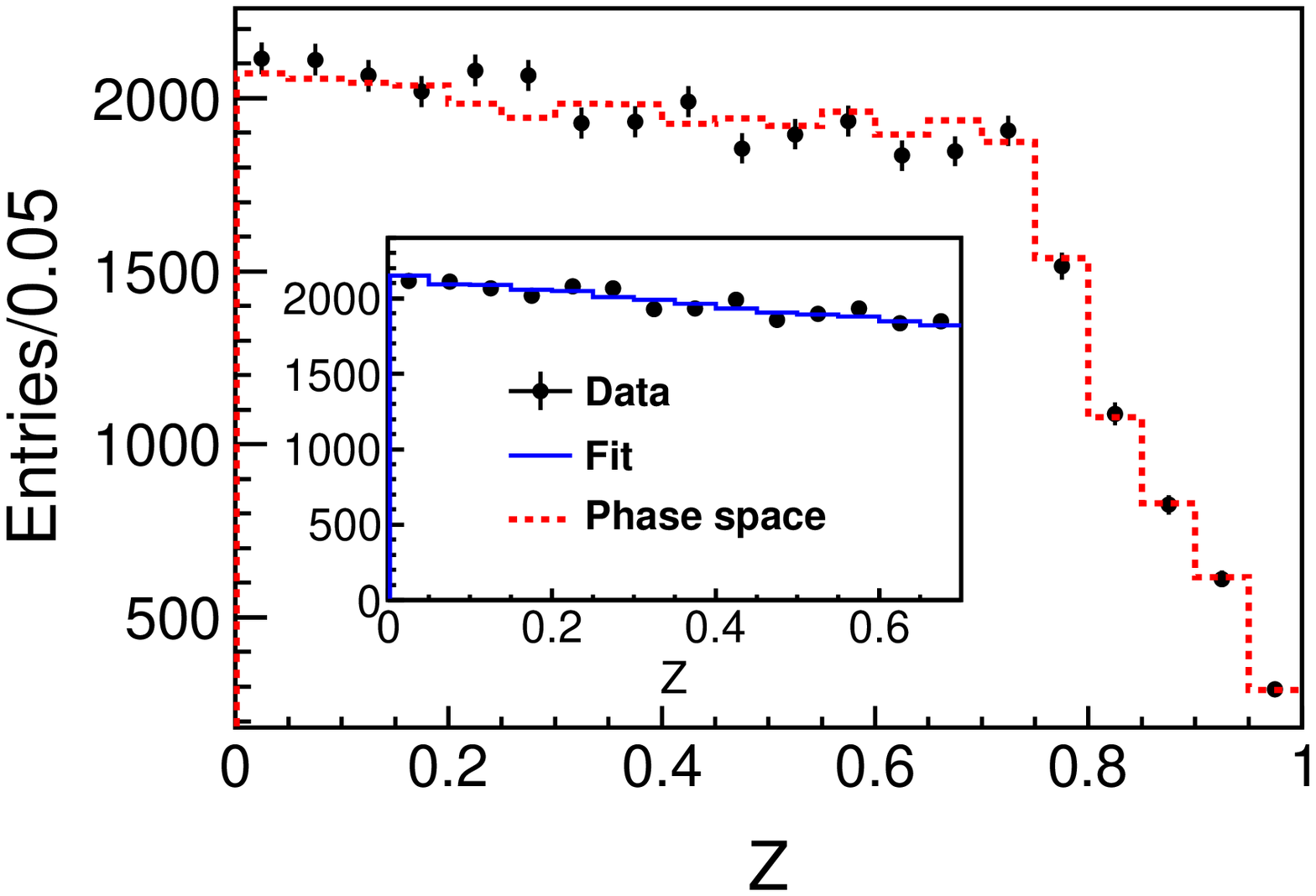}\put(-45,140){\bf (b)}
    \caption{\label{fig:M3pi} (a) Invariant mass spectrum of $\pio\pio\pio$ for $\eta\ra\pio\pio\pio$ obtained
		from data (dots with error bars) and estimated from the inclusive MC sample (shaded histogram).
		(b) Distribution of the kinematic variable $Z$ for $\eta\ra\pio\pio\pio$ obtained from data (dots with error bars)
		and phase space distributed MC events,	where the $Z$ distribution is flat from $Z=0$ to $Z\sim0.76$
		and then drops to zero at $Z=1$ (dashed line).
		The inset shows the $Z$ range which is used for the
		fit to extract the slope parameter $\alpha$. Overlaid on the data is the result of the fit (solid line in the inset).}
\end{figure*}

For $\eta\ra\pio\pio\pio$, the invariant mass spectrum of $\pio\pio\pio$ is shown in Fig.~\ref{fig:M3pi}(a).
A very clean $\eta$ signal is observed. The invariant mass spectrum of $\pio\pio\pio$ obtained from the inclusive MC sample
is also shown, indicating a very low background level of
0.3\% under the $\eta$ signal.
The background is also estimated from the data using $\eta$ sideband regions ($0.49<M(\pio\pio\pio)<0.51$ GeV/$c^2$
and $0.59<M(\pio\pio\pio)<0.61$ GeV/$c^2$),
and is found to be about 1\%. For the determination of the slope parameter $\alpha$, the backgrounds are neglected.

To improve the energy resolution of the $\pio$ candidates and thus the resolution of the Dalitz plot variable $Z$,
the kinematic fit as described above is repeated with the additional constraint that the $\pio\pio\pio$
invariant mass corresponds to the nominal $\eta$ mass.

Finally, a clean sample of 33,908 $\eta\ra\pio\pio\pio$ events is selected.
The distribution of the variable $Z$,
defined in Eq.~(\ref{zdef}), is displayed in Fig.~\ref{fig:M3pi}(b).
The dotted histogram in the same plot represents the MC simulation of phase space events with
$\alpha=0$, as expected at leading order in ChPT.
Due to the kinematic boundaries, the interval of $0<Z<0.7$,
corresponding to the region of phase space in which the $Z$ distribution is flat, is used to extract the slope parameter $\alpha$ from the data.

\begin{figure*}[!htbp]
    \includegraphics[width=8cm, height=6cm]{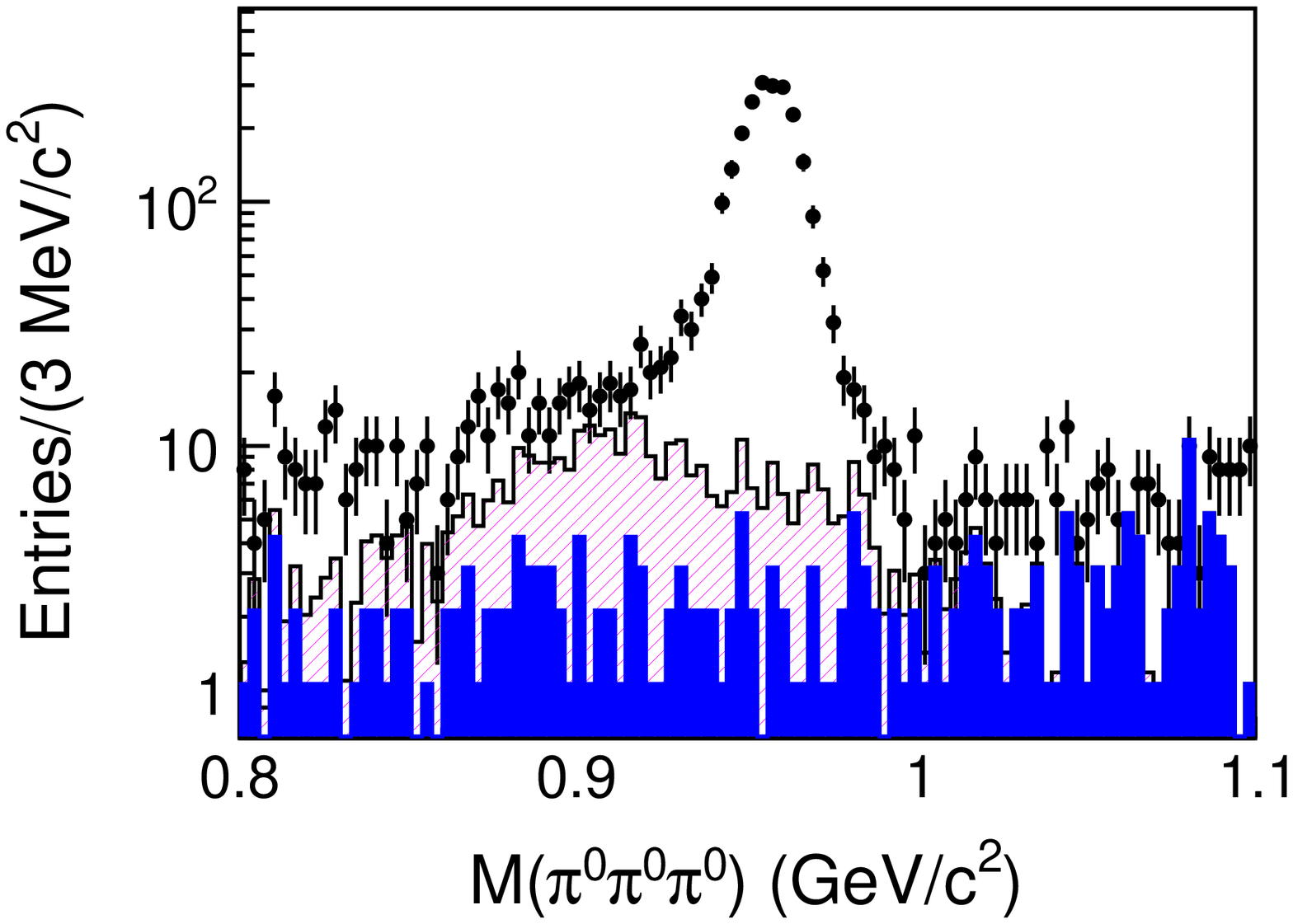}\put(-45,140){\bf (a)}
    \includegraphics[width=8cm, height=6cm]{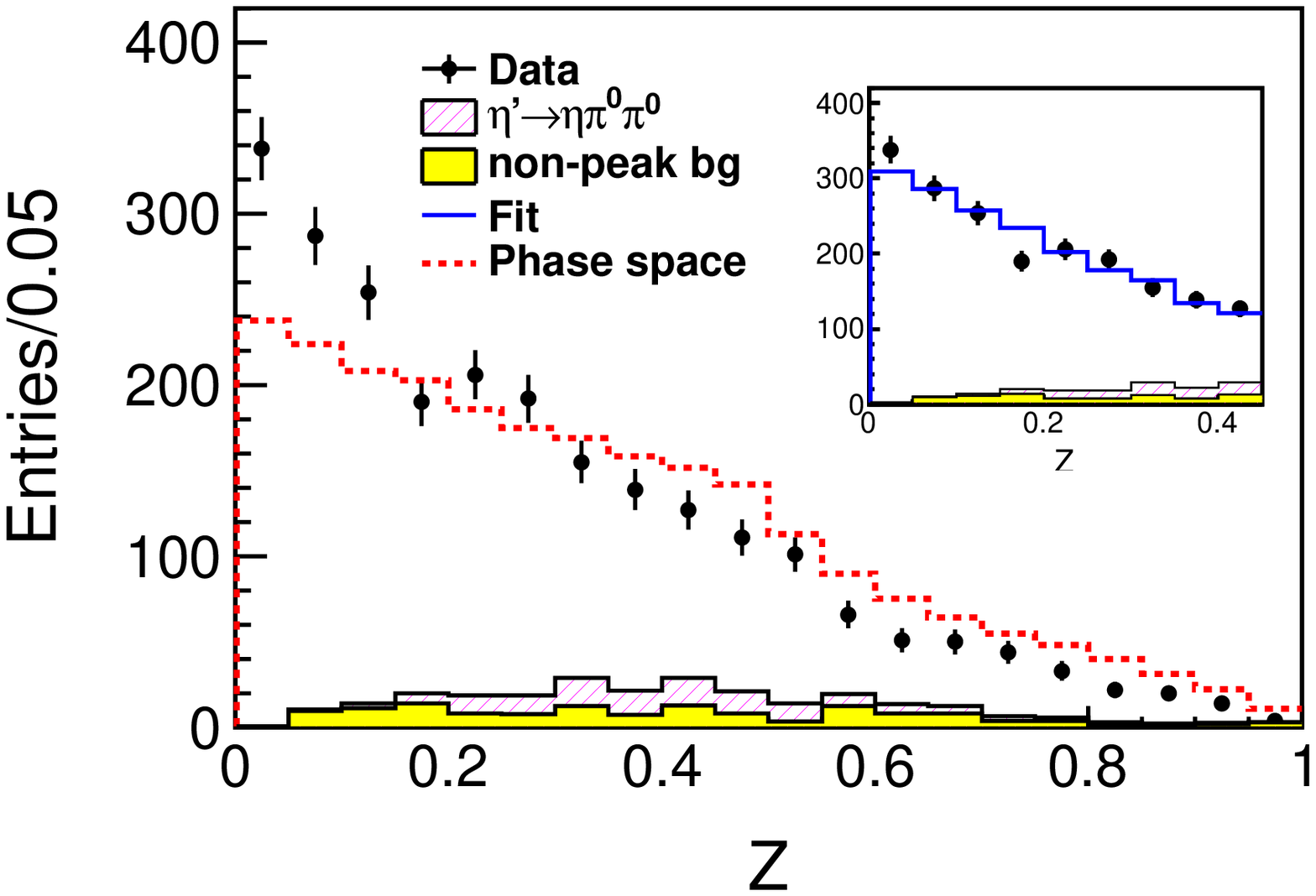}\put(-45,140){\bf (b)}
    \caption{\label{fig:zdistr} (a) Invariant mass spectrum of $\pio\pio\pio$ for $\etap\ra\pio\pio\pio$ obtained
		from the data (dots with error bars), estimated from the inclusive MC sample (shaded),
		and $\etap\ra\eta\pio\pio$ MC events (hatched).
		(b) Distribution of the kinematic variable $Z$ for $\etap\ra\pio\pio\pio$ obtained from the data (dots with error bars),
		phase space distributed MC events
		(dashed line),
		$\etap$ sideband regions (shaded)
		and $\etap\ra\eta\pio\pio$ MC events (hatched). The result of the fit (solid line) is overlaid on the data in the
		insert.}
\end{figure*}

Analogous to the measurement for $\eta\ra\pip\pim\pio$,
an unbinned maximum likelihood fit is performed on the $Z$ distribution of the data to
extract the slope parameter.
The probability density function is constructed with Eq.~(\ref{eq:etaneu_amp})
convoluted with a double Gaussian function
and multiplied by a first-order Chebychev polynomial to account for the resolution $\sigma(Z)$
and detection efficiency $\varepsilon(Z)$, respectively.
Both the resolution and the efficiency functions are obtained from the phase space distributed MC events.
The fit yields $\alpha = -0.055 \pm 0.014$, where
the error is statistical only. In the inset of Fig.~\ref{fig:M3pi}(b) the result of the fit is overlaid
on the distribution for the data.

For $\etap\ra\pio\pio\pio$, the invariant mass spectrum of $\pio\pio\pio$ is shown in Fig.~\ref{fig:zdistr}(a),
where an $\eta^\prime$ signal is clearly visible. The analysis of the $J/\psi$ inclusive decay samples shows that
the dominant background contribution is from $\etap\ra\eta\pio\pio$.
Additional backgrounds are created by $\jpsi$ decays to the same final state,
$e.g.$, $J/\psi\rightarrow\omega\pi^0\pi^0$
with $\omega\rightarrow\gamma\pi^0$. To evaluate the contribution from $\etap\ra\eta\pio\pio$,
$4\times10^6 \jpsi\ra\gamma\etap$ events with $\eta^\prime\rightarrow\eta\pi^0\pi^0$ are generated.
The $\etap$ decay dynamics are modeled according to the results of the Dalitz plot analysis
given in Ref.~\cite{ref:etapipi}.
The invariant mass spectrum of $\pio\pio\pio$ is also shown in Fig.~\ref{fig:zdistr}(a),
where the number of events is scaled to the number of $\jpsi$ events in the data sample,
taking into account the branching fractions of $J/\psi\rightarrow\gamma\eta^\prime$ and the subsequent decays.
Other background contributions ($e.g.$ from $J/\psi\rightarrow\omega\pi^0\pi^0$) are estimated from the data sample using the $\etap$ sideband
regions, defined as $0.845<M(\pi^0\pi^0\pi^0)<0.88$ GeV/$c^2$ and $1.008<M(\pi^0\pi^0\pi^0)< 1.043$ GeV/$c^{2}$
(Fig.~\ref{fig:zdistr}(a)). The total background contamination is estimated to be 11.2\% in the $\etap$ signal mass region
($0.92<M(\pi^0\pi^0\pi^0)<0.99$ GeV/$c^2$). 

After requiring the invariant mass of $\pio\pio\pio$ to be in the $\etap$ signal mass region,
the distribution of $Z$ is shown in Fig.~\ref{fig:zdistr}(b). The MC simulation of phase space events
clearly deviates from the data.
Analogous to $\eta\ra\pio\pio\pio$, the slope parameter $\alpha$ is determined from an unbinned maximum likelihood fit
to data in the range $0<Z<0.45$ with 1,888 events, taking into account the detection efficiency and resolution.
The background estimated from $\etap\ra\eta\pio\pio$ MC events and the $\etap$ sideband regions
is accounted for by subtracting the likelihood for these events from the likelihood
for data. The normalization of background contribution is fixed at its expected intensity.

The fit yields a slope parameter $\alpha=-0.640\pm0.046$, where the error is statistical only.
The result of the fit is overlaid on the $Z$ distribution for the data in the inset of Fig.~\ref{fig:zdistr}(b).

\section{SYSTEMATIC UNCERTAINTIES}

\begin{table*}[htbp]
 \caption{\label{tab:etacha_syserr} Summary of systematic uncertainties for the measurements of the matrix elements
 (all values are given in \%).}
\begin{ruledtabular}
 \begin{tabular}{ccccccccc}
        Source                          & $a$ & $b$ & $d$  & $f$ &
$\alpha(\eta\ra\pio\pio\pio)$ & $\alpha(\etap\ra\pio\pio\pio)$  \\
\hline
        Efficiency parameterization    & 0.6   & 1.7  & 10.4   & 11.7  & 0.4   &  0.1   \\
        Tracking efficiency            & 0.1   & 0.6  & 0.2    & 0.2   &  -    &  -   \\
    		$\pio$ efficiency              & 0.1   & 2.0  & 1.6    & 1.3   & 3.7   &  1.6   \\
        Fit range                      &  -    &  -   & -      & -     & 3.7   &  3.4   \\
        $\pio$ mis-combination         &  -    &  -   & -      & -     & 2.8   &  1.0   \\
    		Background subtraction         &  -    &  -   & -      & -     &  -    &  6.2  \\
        \hline
   			Total                          & 0.7   & 2.7  & 10.5   & 11.8  & 6.1   & 7.3   \\

\end{tabular}
\end{ruledtabular}
\end{table*}

Various sources of systematic uncertainties on the measured Dalitz plot matrix elements have been investigated.
These include uncertainties due to the efficiency parameterization and uncertainties arising from differences
in the tracking and $\pio$ reconstruction between the data and MC samples.
For the measurement of $\alpha$ for $\eta/\etap\ra\pio\pio\pio$,
additional uncertainties due to the fit range and $\pio$ mis-combination are considered.
Uncertainties for $\alpha$ due to the background estimation for $\etap\ra\pio\pio\pio$ are also assigned.
All the above contributions are summarized in Table~\ref{tab:etacha_syserr},
where the total systematic uncertainty is given by the quadratic sum of the individual errors, 
assuming all sources to be independent.
Assuming the correlation factor between each systematic errors is 1, then
the correlation matrix for systematic errors of $\eta\ra\pip\pim\pio$ is
\begin{equation}
\begin{pmatrix}
   & \vline & 		b & 		d &   f\\\hline
 a & \vline & -0.71 &  \phantom{-} 0.99 & -0.97\\
 b & \vline & \phantom{-}  1.00 & -0.73 & \phantom{-}  0.54\\
 d & \vline &       &  \phantom{-} 1.00 & -0.96\\
\end{pmatrix}
.
\end{equation}
In the following, the estimation of the individual uncertainties are discussed in detail.

To estimate the uncertainty due to efficiency parameterizations, we perform alternative fits by changing the
description of the efficiency from polynomial functions to the average efficiencies of local bins.
The change in the obtained values for the matrix elements
from the alternative fits with respect to the default values is assigned as the systematic
uncertainty due to the efficiency parameterization.

Differences between the data and MC samples for the tracking efficiency of charged pions are investigated using
$\jpsi\ra p\bar{p}\pip\pim$ decays. A momentum-dependent correction is obtained for charged pions reconstructed
from MC events. Similarly, a momentum-dependent correction for the $\pio$ efficiency in the MC sample is obtained from $\jpsi\ra\pip\pim\pio$ decays.
The fits to extract the matrix elements are repeated as described above, taking into account the efficiency correction
for charged pions and $\pio$. The change of the matrix elements
with respect to the default fit result is assigned as a systematic uncertainty.

The slope parameter $\alpha$ for $\eta/\etap\ra\pio\pio\pio$ is extracted from a fit to the data
in the kinematic region where the $Z$ distribution of phase space is flat.
By altering the fit range to $0<Z<0.65 (0.68)$ for $\eta\ra\pio\pio\pio$ and $0<Z<0.43 (0.45)$ for $\etap\ra\pio\pio\pio$ and repeating the fit to the data,
the larger changes in $\alpha$ with respect to the default fits are noted and assigned as the systematic uncertainties.

Mis-reconstruction of $\pio$ candidates in true signal events can lead to a wrongly reconstructed position
of the event on the Dalitz plot, and therefore affect the fitted
parameters.
Using signal MC, the possible mis-combination of photons has been studied by matching the generated
photon pairs to the selected $\pio$ candidates. The fraction of events with a mis-combination of photons is 5.4\% 
for $\eta\ra\pio\pio\pio$ and 0.95\% for $\etap\ra\pio\pio\pio$,
respectively. 
Applying the fit to the truth-matched simulated events
only, the impact on the fit parameters is found to be 2.8\% for
$\eta\ra\pio\pio\pio$ and 1.0\% for $\etap\ra\pio\pio\pio$,
respectively.  This is taken as the systematic uncertainty.

In the determination of $\alpha$ for $\etap\ra\pio\pio\pio$, background contributions are
estimated from MC simulations for $\etap\ra\eta\pio\pio$ and $\etap$ sideband regions.
For the peaking background from $\etap\ra\eta\pio\pio$, the uncertainties of the branching fractions
for $\jpsi\ra\gamma\etap$ and $\etap\ra\eta\pio\pio$ taken from Ref.~\cite{PDGgroup} are considered.
In addition, an alternative set of matrix element parameters for $\etap\ra\eta\pio\pio$ as reported
by the GAMS-4$\pi$ collaboration in Ref.~\cite{ref:etapipi} is used in the MC simulation. 
The uncertainty from non-peaking backgrounds is estimated by varying the sideband regions to $0.723<M(\pio\pio\pio)<0.758$ GeV/$c^2$ and
$1.063<M(\pio\pio\pio)<1.098$ GeV/$c^2$. 

In order to estimate the impact from the different resolution of Dalitz plot variables between data and MC sample,
we perform alternative fits
in which the resolution is varied by $\pm$10\% and find that the change of the results is negligible, as expected.

\section{SUMMARY}

Using $1.31\times 10^9$ $\jpsi$ events collected with the BESIII
detector, the Dalitz plots of $\eta\ra\pi^{+}\pi^{-}\pi^{0}$ and $\eta/\etap\ra\pio\pio\pio$
are analyzed and the corresponding matrix elements are extracted.

\begin{figure*}
		\includegraphics[width=0.6\textwidth]{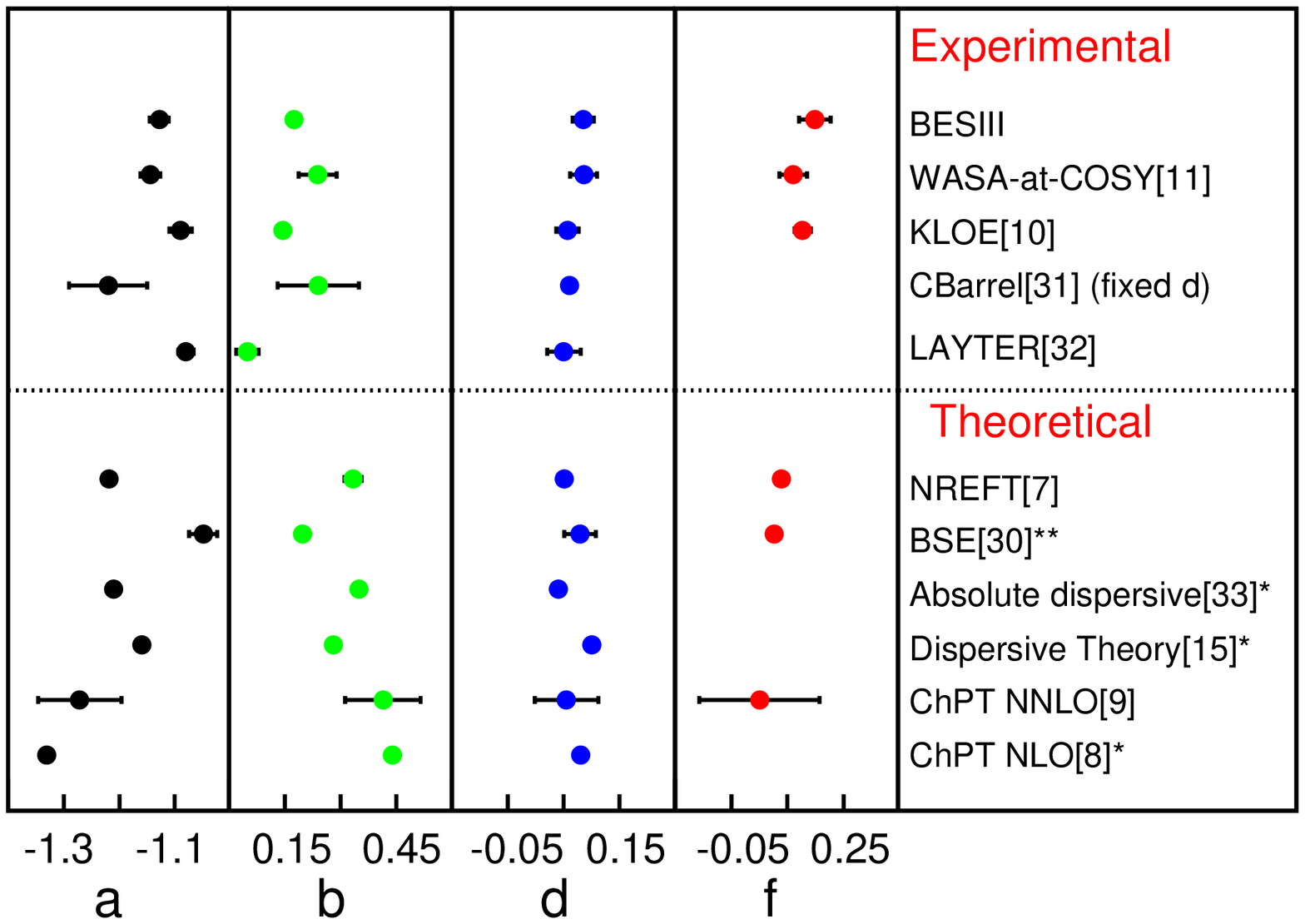}
    \caption{\label{fig:charesult}
	Comparison of experimental measurements and theoretical predictions of the matrix elements for $\eta\ra\pip\pim\pio$.
		$^*$ Theoretical predictions without error.
    $^{**}$BSE denotes Bethe-Salpeter Equation.
		}
\end{figure*}

In the case of charge conjugation invariance,
the Dalitz plot matrix elements for $\eta\rightarrow\pi^+\pi^-\pi^0$ are
determined to be
\begin{displaymath}
\begin{matrix}
a &=&-1.128 \pm 0.015  \pm 0.008,\\
b &=&\phantom{-} 0.153 \pm 0.017  \pm 0.004,\\
d &=&\phantom{-} 0.085 \pm 0.016  \pm 0.009,\\
f &=&\phantom{-} 0.173 \pm 0.028  \pm 0.021,\\
\end{matrix}
\end{displaymath}
where the first errors are statistical and the second ones systematic, here and in the following.
In Fig.~\ref{fig:charesult} our measurement is compared to previous
measurements and theoretical predictions. Our results are in agreement with
the two most recent measurements, and consistent with the predictions
of the dispersive approach and ChPT at NNLO level.

To investigate the charge conjugation violation in
$\eta\rightarrow\pi^+\pi^-\pi^0$, the matrix elements $c$ and $e$ have been determined from a fit to the data.
The obtained values are consistent with zero, while the other
parameters are found to be consistent with those obtained from the fit assuming charge conjugation invariance.
No significant charge symmetry breaking is observed.

\begin{figure*}
		\includegraphics[width=0.48\textwidth]{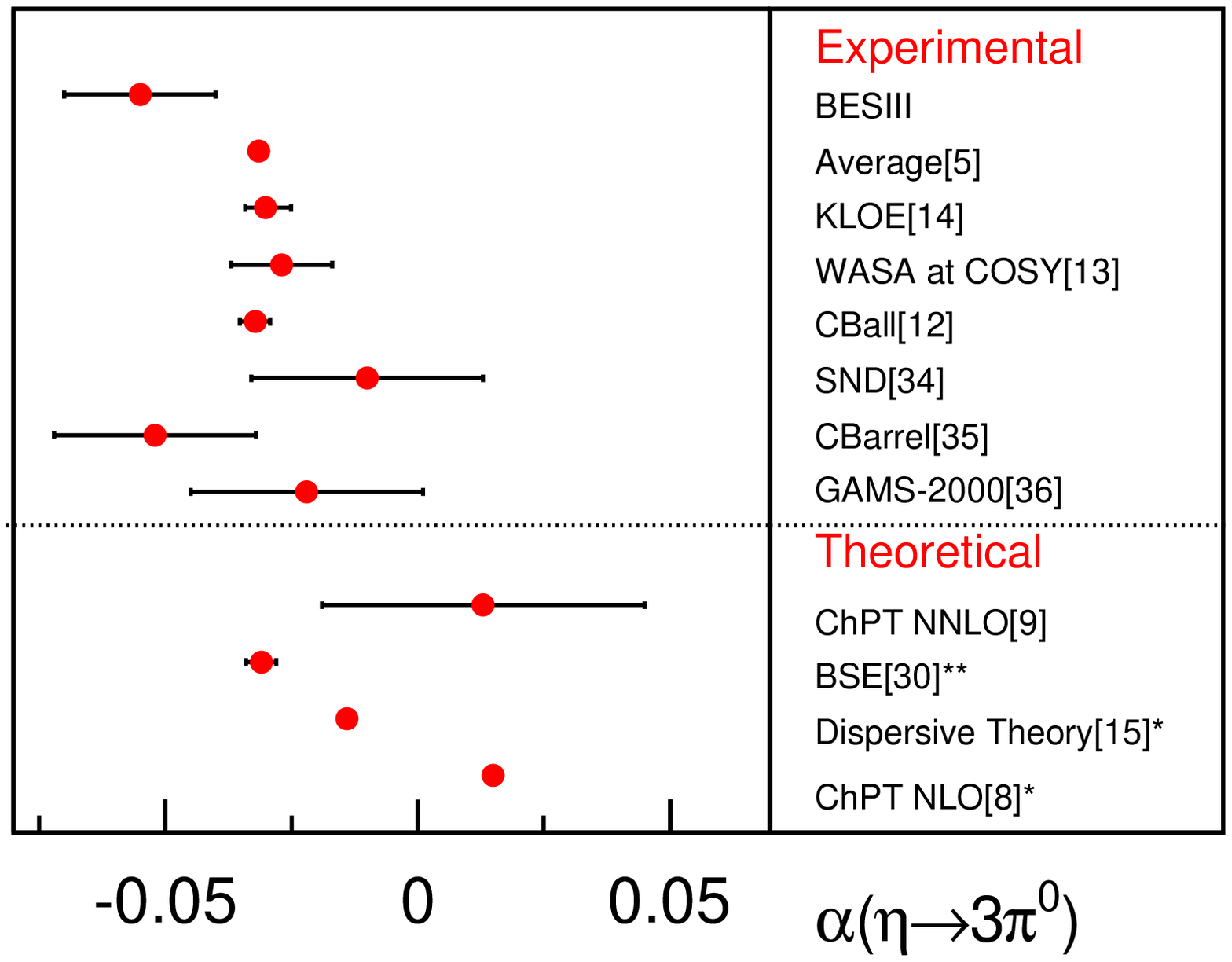}\put(-40,120){\bf (a)}
		\includegraphics[width=0.48\textwidth]{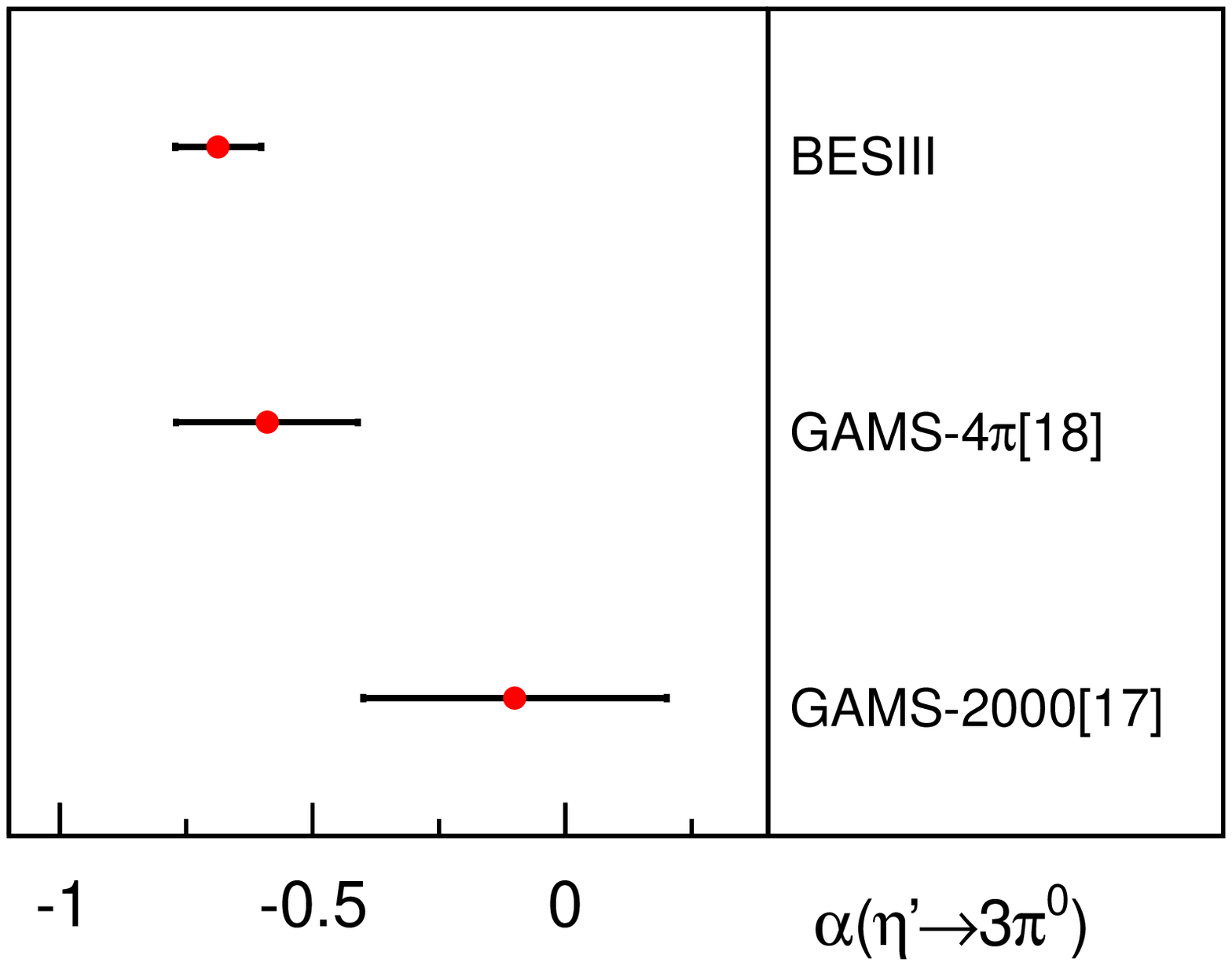}\put(-40,120){\bf (b)}
    \caption{\label{fig:res}
			Comparison of experimental measurements and theoretical predictions of the matrix elements for (a)
                        $\eta\ra\pio\pio\pio$ and (b) $\etap\ra\pio\pio\pio$. 
		$^*$ Theoretical predictions without error. 
		$^{**}$BSE denotes Bethe-Salpeter Equation.
			}
\end{figure*}

After taking into account the systematic uncertainties,
the slope parameter $\alpha$ for $\eta\rightarrow\pi^0\pi^0\pi^0$ is measured to be $-0.055 \pm 0.014 \pm 0.004$.
A comparison to previous works, illustrated in Fig.~\ref{fig:res}(a), indicates that the BESIII result is
compatible with the recent results from other experiments and in agreement with the prediction from ChPT at
NNLO within two standard deviations of the theoretical uncertainties.

The Dalitz plot slope
parameter for $\eta^\prime\rightarrow\pi^0\pi^0\pi^0$ is measured to be
$\alpha = -0.640 \pm 0.046 \pm 0.047$, which is
consistent with but more precise than previous measurements (Fig.~\ref{fig:res}(b)). 
The value deviates significantly from zero.
This implies that final state
interactions play an important role in the decay.
Up to now, there are just a few predictions about the slope parameter of $\etap\ra\pio\pio\pio$.
In Ref.~\cite{CRoiesnel}, the slope parameter is predicted to be less than 0.03, which is excluded by
our measurement.
More recently, using a chiral unitary approach, an expansion of the decay amplitude up to the fifth and sixth
order of $X$ and $Y$ has been used to parameterize the
Dalitz plot of $\etap\ra\pio\pio\pio$~\cite{BBorasoy}. The coefficient, which corresponds to
$\alpha$ in this paper, is found to be in the range between $-2.7$ and $0.1$, consistent with our measurement.

\begin{acknowledgments}
The BESIII collaboration thanks the staff of BEPCII and the IHEP
computing center for their strong support. This work is supported in
part by National Key Basic Research Program of China under Contract
No.~2015CB856700; National Natural Science Foundation of China (NSFC)
under Contracts Nos.~11175189, 11125525, 11235011, 11322544, 11335008, 11425524;
Youth Science Foundation of China under constract No.~Y5118T005C;
the Chinese Academy of Sciences (CAS) Large-Scale Scientific Facility
Program; the CAS Center for Excellence in Particle Physics (CCEPP);
the Collaborative Innovation Center for Particles and Interactions
(CICPI); Joint Large-Scale Scientific Facility Funds of the NSFC and
CAS under Contracts Nos.~11179007, U1232201, U1332201, U1232101; CAS under
Contracts Nos.~KJCX2-YW-N29, KJCX2-YW-N45; 100 Talents Program of CAS;
INPAC and Shanghai Key Laboratory for Particle Physics and Cosmology;
German Research Foundation DFG under Contract No.~Collaborative
Research Center CRC-1044; Istituto Nazionale di Fisica Nucleare,
Italy; Ministry of Development of Turkey under Contract
No.~DPT2006K-120470; Russian Foundation for Basic Research under
Contract No.~14-07-91152; U.S.~Department of Energy under Contracts
Nos.~DE-FG02-04ER41291, DE-FG02-05ER41374, DE-FG02-94ER40823,
DESC0010118; U.S.~National Science Foundation; University of Groningen
(RuG) and the Helmholtzzentrum fuer Schwerionenforschung GmbH (GSI),
Darmstadt; WCU Program of National Research Foundation of Korea under
Contract No.~R32-2008-000-10155-0.

\end{acknowledgments}

\end{document}

%% file: authors_jun2015.tex
\author{
  \begin{small}
    \begin{center}
      M.~Ablikim$^{1}$, M.~N.~Achasov$^{9,f}$, X.~C.~Ai$^{1}$,
      O.~Albayrak$^{5}$, M.~Albrecht$^{4}$, D.~J.~Ambrose$^{44}$,
      A.~Amoroso$^{48A,48C}$, F.~F.~An$^{1}$, Q.~An$^{45,a}$,
      J.~Z.~Bai$^{1}$
% ...
      \\
      \vspace{0.2cm}
      (BESIII Collaboration)\\
      \vspace{0.2cm} {\it
        $^{1}$ Institute of High Energy Physics, Beijing 100049, People's Republic of China\\
% ...
$^{4}$ Bochum Ruhr-University, D-44780 Bochum, Germany\\
        $^{5}$ Carnegie Mellon University, Pittsburgh, Pennsylvania 15213, USA\\
% ...
$^{9}$ G.I. Budker Institute of Nuclear Physics SB RAS (BINP), Novosibirsk 630090, Russia\\
% ...
$^{44}$ University of Rochester, Rochester, New York 14627, USA\\
        $^{45}$ University of Science and Technology of China, Hefei 230026, People's Republic of China\\
% ...
$^{48}$ (A)University of Turin, I-10125, Turin, Italy; (B)University of Eastern Piedmont, I-15121, Alessandria, Italy; (C)INFN, I-10125, Turin, Italy\\
% ...
$^{a}$ Also at State Key Laboratory of Particle Detection and Electronics, Beijing 100049, Hefei 230026, People's Republic of China\\
% ...
$^{f}$ Also at the Novosibirsk State University, Novosibirsk, 630090, Russia\\
% ...
      }
    \end{center}
    \vspace{0.4cm}
  \end{small}
}